\documentclass[10pt]{revtex4-1}
\usepackage{graphicx}
\usepackage{latexsym}
\usepackage{amssymb}
\usepackage{amsmath}
\usepackage{url}
\usepackage{color}

\begin{document}
\newcommand{\crd}{\color {red}}

\def\beq{\begin{equation}}
\def\eeq{\end{equation}}
\def\bsy{\boldsymbol}
\def\av#1{\langle#1\rangle}
\def\etal{{\it et al\/.}}
\def\pc{p_{\rm c}}
\def\l{{\lambda}}
\def\hm{h_*}
\def\xm{x_*}
\def\remark#1{{\bf *** #1 ***}}
\def\ket#1{|#1\rangle}
\def\bra#1{\langle#1|}
\def\braket#1#2{\langle#1|#2\rangle}

\title{Geometric origin of scaling in large traffic networks}

\author{Marko Popovi\'c$^{1,2}$, Hrvoje \v{S}tefan\v{c}i\'c$^2$ \& Vinko
Zlati\'c$^{2,3}$\thanks{vzlatic@irb.hr}}

\affiliation{$^1$ Department of Physics, Faculty of Science, University of Zagreb, P.O.Box
331
HR-10002 Zagreb, Croatia}
\affiliation{$^2$ Theoretical Physics Division, Rudjer Bo\v{s}kovi\'{c} Institute, P.O.Box
180,
HR-10002 Zagreb, Croatia}
\affiliation{$^3$ INFM-CNR Centro SMC Dipartimento di Fisica, 
Sapienza Universit\`a di Roma Piazzale Moro 5, 00185 Roma, Italy}

\begin{abstract}
Large scale traffic networks are an indispensable part of contemporary human
mobility and international trade. Networks of airport travel or cargo ships
movements are invaluable for the understanding of human mobility
patterns\cite{Guimera2005},
epidemic spreading\cite{Colizza2006}, global trade\cite{Imo2006} and spread of invasive
species\cite{Ruiz2000}. Universal features of such networks are necessary ingredients of
their description and can point to important mechanisms of their formation. Different
studies\cite{Barthelemy2010} point to the universal character of some of the exponents
measured in such networks. Here we show that exponents which relate i) the strength of
nodes to their degree and ii) weights of links to degrees of nodes that they connect
have a geometric origin. We present a simple robust model which exhibits the observed
power
laws and relates exponents to the dimensionality of 2D space in which traffic networks are
embedded. The model is studied both analytically and in simulations and the conditions
which result with previously reported exponents are clearly explained. We show that the
relation between weight strength and degree is $s(k)\sim k^{3/2}$, the relation between
distance strength and degree is $s^d(k)\sim k^{3/2}$ and the relation between weight of
link and degrees of linked nodes is $w_{ij}\sim(k_ik_j)^{1/2}$ on the plane 2D surface. We
further analyse the influence of spherical geometry, relevant for the whole planet, on
exact values of these exponents. Our model predicts that these exponents should be found
in future studies of port networks and impose constraints on more refined models of port
networks.
\end{abstract}


\maketitle

An intense empirical research of traffic networks of
aeroplanes\cite{Barrat2004,Guimera2004,Barrat2005,Bagler2008,Zhang2010,Brockman,
Guimera2005,Li2004} and cargo
ships\cite{Hu2009,Kaluza2010,Xu2007} points to the fact that there are exponents of power
laws universally present in these types of networks. In general, nodes in such traffic
networks represent ports. Weighted links of these networks represent fluxes of some
units - people or goods - between these ports\cite{Barrat2004}. In general these fluxes
are directed, but it is also common to represent overall fluxes between ports as
undirected, which is the approach that we follow in this study. The weighted link $w_{ij}$
found in empirical networks therefore represents some average number of
transported objects back and forth, during some time interval between ports $i$ and $j$.
The strength $s_i$ of the port $i$ is
calculated as a sum of fluxes $w_{ij}$ of units through the port i.e. $s_i=\sum_{j}
w_{ij}$. The distance strength of
the port $s^d_i$ is a sum of distances to the neighbouring ports i.e. $s^d_i=\sum_{j\in
n(i)} d_{ij}$,  where $n(i)$ represents the set of ports connected to the port $i$.
Different
studies
\cite{Barrat2004,Barrat2005,Guimera2004,Guimera2005,Li2004,Hu2009,Kaluza2010,Xu2007}
observed interesting relations between aforementioned properties and topological
variables. Namely,
the weighted strength of the port $i$ depends on its number of neighbours $k_i$ as
$s_i\sim
k_i^{\beta}$ \cite{Barrat2004,Guimera2004,Guimera2005,Li2004,Hu2009,Kaluza2010, Xu2007}.
The
distance  strength of the port $i$ scales with its number of neighbours $k_i$ as
$s^d_i\sim
k_i^{\beta_d}$ \cite{Barrat2005}. Finally the flux between two ports of degrees $k_i$ and
$k_j$ scales as $w_{ij}\sim
(k_ik_j)^{\alpha}$ \cite{Barrat2004,Li2004}. Values of these exponents in different
studies
are presented in Table \ref{MeasuredExponents}. Exponents in airport networks have
been much more studied than exponents of cargo ship networks and we do not know for
any attempt of measuring $\alpha$ in cargo ship networks. Nevertheless, agreement
between findings in cargo ship networks and predictions of our model lead us to believe
that our model could capture the behaviour of cargo ship networks as well.
\begin{table}
\centering
\begin{tabular}{|p{7 cm}|| p{2 cm}| p{1 cm}| p{1.5 cm}|}
\hline
Study & $\beta$ & $\beta^d$ & $\alpha$\\
\hline\hline
Barrat \etal \cite{Barrat2004} World Airport Network (WAN)& $1.5\pm 0.1$ &  & $0.5\pm
0.1$\\
\hline
Barrat \etal \cite{Barrat2005} North American Airport Network & $1.7$ & $1.4$ & \\
\hline
Bagler \cite{Bagler2008} Airport Network of India& $1.43\pm 0.06$ & & \\
\hline
Zhang \etal\cite{Zhang2010} & $1.58$ & & \\
Passenger Airport Network of China & & & \\
\hline
Zhang \etal\cite{Zhang2010} & $2.2$ & & \\
Cargo Airport Network of China & & & \\
\hline
O'Danleyman \etal\cite{Brockman} (WAN) & $1.33$ & & \\
\hline\hline
Kaluza \etal \cite{Kaluza2010} Network of Global Cargo Ship Movements (NGCSM) & $1.46 \pm
0.1$ & &
\\
\hline
Hu and Zhu\cite{Hu2009} World Wide Maritime Transportation Network & $1.3$ & & \\
\hline
O'Danleyman \etal\cite{Brockman} (NGCSM)& $1.33$ & & \\
\hline \hline
\end{tabular}
\caption{\label{MeasuredExponents}Measured exponents in different studies. Except for the
exponent $\beta$ of Cargo Airport Network of China, all the reported exponents fit well
to the predictions of our model. Li and Cai\cite{Li2004} also suggested the value of
$\alpha=1/2$ for the Chinese Airport Network as an ansatz for their data, although without
clear fitting procedure. Most of the studies did not report error intervals for the
measured
exponents but, based on those that did, we can assume errors of around $0.1$. }
\label{tab:analytical}
\end{table}


Although a number of models coupling topological and traffic
properties of such spatial networks has been proposed, none of these models managed to
explain
all three observed exponents. In\cite{Mukherjee2006} authors proposed a model based on
spatial preferential attachment\cite{Barabasi1999}. They found a relation between
exponents, which explains only one or the other observed exponent, but
not all of them. Further, in\cite{Barrat2004Prl,Barrat2005} authors studied the
model of weight driven attachment which was able to explain some of the observed
properties of transport networks, but failed to reproduce exponents which relate degree
properties with weight properties. Their model yields the value of $\beta=1$, and the
authors proposed that the measured exponents are a consequence of some non-linear process
lying behind the attachment model they have used. The most complete analysis of coupling
between topology and weights is laid out in\cite{Ginestra2005}, in which the author
proposed two models - a weighted attachment model with the addition of weights and a
fitness based
model\cite{Caldarelli2002}. These models can reproduce any possible
value of $\beta$, including $\beta=3/2$, and a number of other network properties, but are
sensitive to tunable model parameters, and exponents $\beta^d$ and $\alpha$ were not
reproduced.

We observe that values of measured exponents point to the possibility that they are
rational
numbers. Traditional statistical mechanics\cite{Goldenfled1992} suggest that rational
exponents can often be attributed to the dimensionality of space. If the values of
measured exponents are really of geometric origin, then a fairly general model should be
able to reproduce these exponents. Such a model should produce relatively stable exponents
 with respect to other model parameters and possible additional refinements of the
model. 

We propose a simple and robust mathematical model of port networks. This model is based on
three assumptions. First, we assume that the probability for an individual unit to travel
from one port to the other is determined by some hidden variables assigned to all
ports. These variables depend on properties of the port such as the 
size of population, its economic power, number of international companies, etc., but
could also depend on the traffic network funcionality to include transient ports etc.
The predictions of the model do not depend on the precise definition of these
variables or the precise mechanism of their assignement to ports. These
hidden variables are modelled as the fitness of every port - a random number $x$, which is
attached to every port and is drawn from some probability distribution
$\rho(x)$\cite{Caldarelli2002}. We will later show that exponents are very robust to the
choice of probability distribution and to possible spatial correlations between different
fitness variables. The second assumption is that the probability of intended travel
between
two
ports is proportional to the product of their fitness variables $p(i\rightarrow j)\sim
x_ix_j$. This assumption is related to the usual model of random mixing in
networks\cite{Newman2003}. The fitness variables represent some units which travel back
and
forth to interact with some other units. The probability that some unit from port $i$ will
travel to interact with some unit from port $j$ is, in its simplest instance, proportional
to the amount of units in port $j$. This assumption means that the expected flux $w_{ij}$
of units
from port $i$ to port $j$ and from port $j$ to port $i$ is proportional to the product of
fitness variables $w_{ij}\sim x_ix_j$. The third assumption is related to the economic
and geometric properties of traffic networks. We assume that the link between ports $i$
and $j$ will exist only if the expected volume of traffic will cover the costs of distant
travel.
The costs of travel $c_{ij}$ from port $i$ to port $j$ are modelled as a linearly
increasing
function of their distance $r_{ij}$. This
assumption is related to the fact that costs of travel are, among other things, determined
by the amount of fuel one has to spend travelling from one place to another and by the
costs of crews which are also proportional to the distance via the number of travels a
crew can
manage in the unit of time. We also present the data on the travel fares between
different U.S. airports and their mutual distances\cite{AirFareData}. The relationship
between distance and fares has Pearson's correlation $r=0.64$ (Supplementary Figure
1), which justifies our assumption of linearity in the model. Notice that this data state
real
commercial fares which are not equal to
the lower bound of transport profitability, which is relevant for this model.
Nevertheless,
the linear relationship between fares and distance is still clearly
present. We expect that this relationship would be even more pronounced in the data on
pure costs of travel between different ports.


The proposed model is studied both analytically and in simulations. Let us first
consider an idealized case in which ports are randomly spread on an infinite 2-dimensional
plane. We assume that the distribution of ports is relatively homogeneous and model their
spatial distribution as a spatial Poisson process\cite{Feller}. To every existing port $i$
a random number $X(i)$ drawn from some probability distribution $\rho(x)$ is assigned.
Then a relation between
the expected degree $\langle k(X)\rangle$ of the port and its fitness variable $X$ can be
written as
\begin{equation}\label{Eq: k}
\langle k(X)\rangle = \sigma\int
dA\int_{0}^{\infty}\Theta\left(f(xX)-c(r)\right)\rho(x)dx.
\end{equation}
Here $\sigma$ is a surface density of ports, $dA$ is a differential element of the
surface, $\Theta$ is a Heaviside step function, $f$ is function of earnings generated from
the flux of travelling units, and $c$ is a cost function of travel which grows with
distance. We assume that these functions are monotonously increasing. In the rest of this
paper we assume $f(u)=c(u)=u$ as the simplest
linear relationship. In Supplementary Information 1 we give a short analysis of some other
possibilities for functions $f$ and $c$ which lead to the same exponents. In it, we also
show that the relation $w_{ij}=x_ix_j$ needs to be satisfied on the average only to
produce
the same exponents as the exact version of the model.

Similarly to equation (\ref{Eq: k}) the expected strength and expected distance strength
of the port can be related to its fitness $X$ as
\begin{eqnarray}
\langle s(X)\rangle=\sigma \int
dA\int_{0}^{\infty}xX\Theta\left(xX-r\right)\rho(x)dx.\\
\langle s^d(X)\rangle=\sigma \int
dA\int_{0}^{\infty}r\Theta\left(xX-r\right)\rho(x)dx.
\end{eqnarray}
The integrations in these equations are easily performed to obtain $\langle
k(X)\rangle=\pi \sigma M_2(\rho)X^2$,
$\langle s(X)\rangle=\pi \sigma M_3(\rho) X^3$ and $\langle s^d(X)\rangle=\frac{2}{3}\pi
\sigma M_3(\rho) X^3 $. Here $M_{2,3}(\rho)$ represent the second and the third moment of
the
distribution $\rho(x)$. From these relations immediately follows that $\langle
s(k)\rangle\sim k^{3/2}$, $\langle s^d(k)\rangle\sim k^{3/2}$ and $
\langle w(k_i,k_j)\rangle\sim (\langle k_i\rangle \langle k_j\rangle)^{1/2}$. For
simplicity we refer to these exponents as
infinite exponents in the remaining text. One can also understand this relationship
qualitatively. Port $i$ will be connected with some port $j$ up to certain radius
$r^c_{ij}$ given by fitness values of these ports $X(i)$ and $X(j)$. Some port $i'$ which
has the double fitness of the port $i$, $X(i')=2X(i)$ will connect with port $j$ on double
distance $r^c_{i'j}=2r^c_{ij}$. Since we also assume the homogeneous distribution of
ports, the number of ports to which port $i'$ is connected grows with the surface i.e.
$k\sim r^2 \sim
X^2$. Weights are assumed to be proportional to the fitness $X$ and to the number of
connections $k$ which yields $s\sim Xr^2 \sim X^3$. The average distance of the port is
also
proportional to the distance and to the typical number of
connections $s^d\sim rk \sim X^3$. Having this in mind, it is clear that the relationship
between the cut off distance $r^c$ and the fitness $X$ has to be linear in order to
reproduce
measured exponents. 

Although distribution $\rho(x)$ can, in principle, be freely
chosen on the infinite 2D plane, it is constrained for the applications by the observed
degree distributions
$p_o(k_o)$ extracted from the data. If we assume that $k_o(x)$ is well approximated by
$\langle k(x)\rangle$ and since the relationship between expected degree $\langle
k(x)\rangle$ and the fitness
variable $x$ is a monotonic function $k=\mathcal{F}(x)\sim x^2$, then the relation 
$\rho(x)\sim|\frac{d\mathcal{F}(x)}{dx}|p_o(\mathcal{F}
(x))$ follows. In the case of scale free distributions $p_o(k)\sim k^{-\gamma}$, the
relation yields $\rho(x)\sim x^{1-2\gamma}$. Relevant
studies\cite{Barrat2004,Bagler2008,Hu2009,Xu2007}
have found power law distributions of degrees in port networks and in validations of
our model we investigate power law distributions of fitnesses for exponents $\gamma \in
\langle 1,5]$. Another constraint on the distribution is the requirement
that every port should be reachable from every other port. In other words, we demand the
existence of a giant component in the network\cite{CaldarelliBook}. Yet another constraint
is that the network should be sparse enough as the real port networks are. This
condition can  be controlled by monitoring the average degree of the network. All these
factors have to be taken in account to choose model parameters, since not every
distribution of fitnesses will satisfy all constraints for all densities or for all
functions $f$ and $c$. These constraints were accounted for by considering the minimal and
maximal values of fitness variables.

In Figure 1 we show dependences of weighted properties on topological properties for
finite 2D space depending on the exponent $\gamma$ of the power law distribution
$\rho(x)$. For a broad range of power law exponents simulated exponents $\alpha$, $\beta$
and $\beta^d$ are close to
theoretically predicted values, as can be seen in Figure 1. The increase of the observed
exponents
as power law is becoming steeper which is related to the process of network
sparsification, as
shown in Figure \ref{Figure2}. In the limiting case of a network consisting of
disconnected
subgraphs it is hard to talk about exponents at all. The degree distribution and
the aforementioned exponents are clearly not the only properties of port networks
previously reported.
In Supplementary Information 2, for the reasons of completeness, we also present
an analytical calculation for the clustering coefficient in infinite 2D plane. For
thelarge $k$ we
observe behavior close to $C(k)\sim k^{-1}$, a feature previously reported by other
researchers
\cite{Barrat2004,Barrat2005}. 

In the Supplementary Information 3 we present an analytical treatment of the
model for the finite 2D plane and for the sphere of radius $R$. The case of the sphere is
particularly important since it represents the real geometry in which port networks
operate. The expected degree $\langle k(X)\rangle$, the strength $\langle s(X)\rangle$ and
the distance strength $\langle s^d(X)\rangle$ of the port with fitness $X$ on a sphere can
be
computed from relations
\begin{equation}
 \langle k(X) \rangle = \sigma R^2 \int\limits_0^{2\pi}d\varphi\int\limits_0^\pi
\sin{\theta}d\theta \int\limits_{R\theta/X}^\infty\rho(x)dx,
\end{equation}
\begin{equation}
 \langle s(X) \rangle = \sigma R^2 \int\limits_0^{2\pi}d\varphi\int\limits_0^\pi
\sin{\theta}d\theta \int\limits_{R\theta/X}^\infty X x \rho(x)dx,
\end{equation}
and
\begin{equation}
 \langle s^d(X)\rangle = \sigma R^2 \int\limits_0^{2\pi}d\varphi\int\limits_0^\pi
\sin{\theta}d\theta \int\limits_{R\theta/X}^\infty \theta R \rho(x)dx.
\end{equation}
The change of the geometry influences the dependences of these variable on
the fitness $X$, and we find deviations from the infinite exponents. Nevertheless, the
average
distances between ports reported in the previous studies, point to the fact that the
majority of the traffic takes place on the locally almost flat elements of a
sphere. The average distance $\langle d\rangle$ of travels in the case of the North
American
airport network is around 1000 km\cite{Barrat2005}, which means that the relative
difference
between $\sin(\langle d\rangle/R)$ and $\langle d\rangle/R$ is only around $0.4\%$.
We believe that this  is the main reason why the monitored exponents of port networks
really resemble theoretical values calculated for the infinite 2D plane.

In real traffic networks, ports are not homogeneously distributed. In the
previous text we dealt with density of ports which are scattered over the space as
Possionian random variables\cite{Feller}. To check if the heterogeneities which
necessarily exist among the real ports influence the behaviour of exponents, we gathered
the
data on geographical locations in the U.S.\cite{GeoLocations}. In Figure
\ref{Figure2}
we present simulations of our model on the geographical locations of American airports.

In Figure \ref{Figure3}, we present three different sets of data points taken from the
database\cite{GeoLocations} and construct a network among them. The exponents do not
differ
much among these different spatial port distributions, as long as the distribution of
the fitness
variable fulfils the aforementioned constraints. The robustness of this model to
the heterogeneity of
spatial port distribution is clear. It is also clear that many ports in the world are used
just as transit airports for a significant amount of traffic and the fitness variables
used
in the model do not capture such behaviour. On the other hand, such usage of airports also
influences the realistic possible values of fitnesses and we expect that in reality they
have a certain spatial hierarchical distribution. 

In conclusion, in this paper we presented a very robust model which can be altered in
many conceivable ways, which generally reproduce observed exponents, which relate
weighted properties to topological properties. Even further it can be shown that in the
case of general D-dimensional flat space, the exponents would be $\alpha=1/D$ and
$\beta=\beta^d=(D+1)/D$ (see Supplementary Information 4), which could (apart from
intellectual curiosity) possibly
be interesting for the case of interstellar traffic with D=3.
Similar exponents of geometric origin have been found in the case of allometric scaling in
biological systems
\cite{West1997}, and at present it is not clear if the presented model is related to these
phenomena.
Furthermore, we believe that some other modes of transportation like, for
instance, intercity buses, could also be well described with this model, but at present we
are not aware of any additional research in that direction.  
Since a convincing simple relationship between fitness variables
proposed in this paper and measurable socio-economic quantities is still lacking, 
further research in this direction is clearly needed.

\section{Methods}
All simulations of the model were performed in Fortran90. The locations
of the ports were drawn randomly from uniform distribution defined on the circle of area
$10$ in arbitrary units in the case of a finite 2D plane. We used different truncated
power-law distribution for fitness variables. The parameters of the distribution were:
minimal value of fitness $X_{min}=0.1$, maximal value of fitness $X_{max}=1.6$ and
exponent $1<\gamma<5$. For every set of points or different parameters we have performed
$100$ different simulations with stochastic variables drawn from power-law distributions.
In the case of the sphere with radius $R=1$ we have used exactly the same parameters.
The locations of ports on the real maps were acquired from geographical data. We have
used geolocations of of airports, hospitals and capes in the U.S. \cite{GeoLocations} in
order to encapsulate as many of real spatial nonpoissonian heterogeneous point
distributions as possible. For the Geolocations we have used $X_{min}=0.01$ or
$X_{min}=0.02$ and $X_{max}=0.65$. We have also used $20$ realizations, due to a large
number of data points. The exponents were calculated by the least square fitting procedure
of the logarithms of related variables. Reported errors are standard deviations of fitted
exponents in the simulation sample. 

\section*{Acknowledgements}

H.S. and V.Z. have been funded through Croatian ministry of education, science and sport
project No. 098-0352828-2863. V.Z. acknowledges partial support from EU FET Open
Project FOC nr 255987. Authors would like to thank G. Caldarelli and A. Gabrielli
for useful comments and to D. Glavan on a very useful comment.

\section*{Contributions}
M.P. developed analytical model and performed all of the simulations and most of the
analytical calculations. H.S. contributed to the development of the model and made a
contribution to analytical treatment. V.Z. conceived the problem, proposed the model, made
a minor contribution to analytical description, created figures and was principal writer
of the paper. All of the authors contributed in the design of the research and
interpretation of the results.
\subsection{Corresponding Authors}
Vinko Zlatic \url{vzlatic@irb.hr}

\newpage

\section*{Supplementary Information 1}
\label{fANDc}

The cost function is generally defined as $f(x_1x_2)-c(r_{12})$. The expected degree of
node
with fitness $X$ is 
\begin{eqnarray}\label{EQ:fANDc}
 \langle k(X)\rangle&=& \sigma\int
dA\int_{0}^{\infty}\Theta\left(f(xX)-c(r)\right)\rho(x)dx\nonumber\\
&=&2\pi\sigma\int_{0}^{\infty}\rho(x)dx\int_{0}^{c^{-1}(f(xX))}rdr\nonumber\\
&=&\pi\sigma\int_{0}^{\infty}[c^{-1}(f(xX))]^2\rho(x)dx.
\end{eqnarray}
Similarly the expected strength of the node with fitness $X$ is
\begin{equation}\label{EQ:fANDc2}
  \langle s(X)\rangle=\pi\sigma\int_{0}^{\infty}[c^{-1}(f(xX))]^2(xX)\rho(x)dx.
\end{equation}
 Equations (\ref{EQ:fANDc}) and (\ref{EQ:fANDc2}) again give the same exponent behaviour
if we
choose $f=c$
\begin{eqnarray}\label{EQ:fANDc3}
\langle k(X)\rangle&=&\pi\sigma\int_{0}^{\infty}[f^{-1}(f(xX))]^2\rho(x)dx\nonumber\\
&=&\pi\sigma X^2M_2(\rho).\\
\langle s(X)\rangle&=&\pi\sigma\int_{0}^{\infty}[f^{-1}(f(xX))]^2(xX)\rho(x)dx\nonumber\\
&=&\pi\sigma X^3M_3(\rho).
\end{eqnarray}
In the case of different functions we can again easily reconstruct proposed exponents. For
instance let  $f(x)=fx$, and $c(x)=c_1x+c_2$. Then equations for the $s$ and $k$ are:
\begin{eqnarray}
 k(X)&=&\pi\sigma\left[\frac{f^2X^2M_2(\rho)}{c_1^2}-2\frac{fc_2XM_1(\rho)}{c_1^2}+\frac{
c_2 ^2}{c_1^2}\right],\nonumber\\
s(X)&=&\pi\sigma\left[\frac{f^2X^3M_3(\rho)}{c_1^2}-2\frac{fc_2X^2M_2(\rho)}{c_1^2}
+\frac {
c_2 ^2XM_1(\rho)}{c_1^2}\right].\nonumber\\
\end{eqnarray}
Proposed exponents will be reconstructed as long as
$\left|\frac{f^2X^2M_2(\rho)}{2fc_2XM_1(\rho)-
c_2 ^2}\right|\ll1$ and
$\left|\frac{f^2X^3M_3(\rho)}{2fc_2X^2M_2(\rho)
- c_2 ^2XM_1(\rho)}\right|\ll1$. In practice it will workj allways when
$fX\langle X\rangle\gg c_2$.

Let us further consider the case in which flow $w_{ij}$ is not necessarily exactly equal
to
the product $x_ix_j$. Realistically, one can expect that some ports are better connected
than expected by chance, while some others are not so well connected. For
example,
airports of two neighbouring states which do not have diplomatic relationships because of
political reasons can have much lower traffic than one would expect from fits of their
overall traffic. This additional preference of the links can be modelled with a new random
variable $\eta$ taken from some distribution $p(\eta)$. Let us assume that the average
value $\langle \eta \rangle=1$. Traffic flows can be written as $w_{ij}=\eta_{ij}x_ix_j$.
If $\eta$ and $x$ are independent random numbers, the equations for degree, strength and
strength distance for infinite 2D plane are $\langle k(X)\rangle=\pi\sigma
M_2(\rho)M_2(p)X^2$, $\langle s(X)\rangle=\pi\sigma
M_3(\rho)M_3(p)X^3$ and $\langle s(X)^d\rangle=2\pi\sigma
M_3(\rho)M_3(p)X^3/3$. Obviously, infinite exponents are preserved even if we allow some
fluctuations around expected values of flows.

\newpage

\section*{Supplementray Information 2}
The expected degree dependent clustering $\langle C(k)\rangle$ coefficient of some port 
on an infinite 2D plane can be calculated as:
\begin{equation}
\langle C(X)\rangle = \frac{\sigma^2 \int\limits_0^\infty r_1 d r_1\int\limits_0^\infty
r_2 d r_2 \int\limits_0^{2\pi}d \varphi_1 \int\limits_0^{2\pi}
d\varphi_2\int\limits_0^\infty\rho(x_1)dx_1\int\limits_0^\infty\rho(x_2)dx_2\Theta(Xx_1-r_
1)\Theta(Xx_2-r_2)\Theta(x_1x_2-r_{12})}{\sigma^2\int\limits_0^\infty r_1 d
r_1\int\limits_0^\infty r_2 d r_2 \int\limits_0^{2\pi}d \varphi_1 \int\limits_0^{2\pi}
d\varphi_2\int\limits_0^\infty\rho(x_1)dx_1\int\limits_0^\infty\rho(x_2)dx_2\Theta(Xx_1-r_
1)\Theta(Xx_2-r_2)}
\end{equation}
where $r_{12}=\sqrt{r_1^2+r_2^2-2r_1r_2\cos(\varphi_1-\varphi_2)}$\\
The denominator $D$ is equal to:
\begin{eqnarray}
D &=&  4 \sigma^2
\pi^2\int\limits_0^\infty\rho(x_1)dx_1\int\limits_0^{Xx_1}
r_1dp_1\int\limits_0^\infty\rho(x_2)dx_2\int\limits_0^{Xx_2}r_2dr_2\\
&=& 4 \sigma^2\pi^2\left[\frac{1}{2}X^2M_2(\rho)\right]^2
\end{eqnarray}
and therefore
\begin{equation}
D = \sigma^2 \pi^2 X^4 \left[M_2(\rho)\right]^2
\end{equation}
The numerator $A$ is equal to
\begin{equation}
A = \sigma^2\int\limits_0^\infty\rho(x_1)dx_1\int\limits_0^{Xx_1} r_1 d
r_1\int\limits_0^\infty\rho(x_2)dx_2\int\limits_0^{Xx_2} r_2 d r_2 \int\limits_0^{2\pi}d
\varphi_1 \int\limits_0^{2\pi} d\varphi_2 \Theta(x_1x_2-r_{12}).
\end{equation}
We can make a convenient substitution
\begin{eqnarray}
\varphi &=& \frac{1}{2}(\varphi_2-\varphi_1)\\
\varphi' &=& \frac{1}{2}(\varphi_2+\varphi_1).
\end{eqnarray}
Separating the integral domain in two parts we can write integrals over new angles:
\begin{equation}\label{domainSep}
\int\limits_{-\pi}^0
d\varphi\int\limits_{-\varphi}^{2\pi+\varphi}d\varphi'\frac{\partial(\varphi_2,\varphi_1)}
{\partial(\varphi,\varphi')}\Theta(x_1x_2-r_{12}) + \int\limits_0^\pi
d\varphi\int\limits_{\varphi}^{2\pi-\varphi}d\varphi'\frac{\partial(\varphi_2,\varphi_1)}
{\partial(\varphi,\varphi')}\Theta(x_1x_2-r_{12}).
\end{equation}
Using Eq. (\ref{domainSep})the numerator can be written as
\begin{equation}
A = \sigma^2\int\limits_0^\infty\rho(x_1)dx_1\int\limits_0^{Xx_1} r_1 d
r_1\int\limits_0^\infty\rho(x_2)dx_2\int\limits_0^{Xx_2} r_2 d r_2 2
\int\limits_0^{2\pi}d\varphi(2\pi-\varphi)\Theta(x_1x_2-r_{12})
\end{equation}
The random variables are positive semidefinite and thus

\begin{equation}
\Theta(x_1x_2-\sqrt{r_1^2+r_2^2-2r_1r_2\cos{\varphi}})=\Theta(\cos{\varphi}-\frac{
r_1^2+r_2^2-(x_1x_2)^2}{2 r_1 r_2}).
\end{equation}
We can write
\begin{equation}
A = 4 \pi^2 \int\limits_0^\infty\rho(x_1)dx_1\int\limits_0^{Xx_1} r_1 d
r_1\int\limits_0^\infty\rho(x_2)dx_2\int\limits_0^{Xx_2} r_2 d r_2
~~\chi\left(\frac{r_1^2+r_2^2-(x_1x_2)^2}{2 r_1 r_2}\right).
\end{equation}
Function $\chi(x)$ is defined as
\begin{equation}
\chi(x) = \begin{cases}
					1 & x < -1\\
					\frac{1}{\pi}\arccos{x} & -1 < x < 1\\
					0 & x > 1
					\end{cases}
\end{equation}
Finally the clustering coefficient is
\begin{equation}
\langle C(X)\rangle =
\frac{4}{X^4\left[M_2(\rho)\right]^2}\int\limits_0^\infty\rho(x_1)dx_1\int\limits_0^{Xx_1}
r_1 d r_1\int\limits_0^\infty\rho(x_2)dx_2\int\limits_0^{Xx_2} r_2 d r_2
~~\chi\left(\frac{r_1^2+r_2^2-(x_1x_2)^2}{2 r_1 r_2}\right)
\end{equation}
Behavior of the clustering coefficient with respect to $k$ is presented in supplementary
figure 2.

\newpage

\section*{Supplementary Information 3}
\label{FiniteAndSphere}
Let us consider a point on the pole of the sphere of radius $R$. The sphere is covered
with random points at locations $\theta$ and $\phi$. The density of these points is given
with $\sigma=\sum_i\int_{\mathcal{S}}\delta(\vec{r}-\vec{r}_i)d\vec{r}/S$. The distance of
points from the pole is given with $R\theta$, where $R$ is  the radius of the sphere and
$\theta$ is the inclination angle. We assume that the point on the pole is a
port with the variable $X$ and that other points represent different ports whose fitnesses
are
drawn from some unspecified distribution $\rho(x)$. The expected degree of the port on
the
pole is then

\begin{equation}
 \langle k(X) \rangle = \sigma R^2 \int\limits_0^{2\pi}d\varphi\int\limits_0^\pi
\sin{\theta}d\theta \int\limits_{R\theta/X}^\infty\rho(x)dx.
\end{equation}
Similarly the strength $\langle s\rangle$ and distance strength $\langle s^d\rangle$ of
the vertex can be calculated from:
\begin{equation}
 \langle s(X) \rangle = \sigma R^2 \int\limits_0^{2\pi}d\varphi\int\limits_0^\pi
\sin{\theta}d\theta \int\limits_{R\theta/X}^\infty X x \rho(x)dx,
\end{equation}
\begin{equation}
 \langle s^d(X)\rangle = \sigma R^2 \int\limits_0^{2\pi}d\varphi\int\limits_0^\pi
\sin{\theta}d\theta \int\limits_{R\theta/X}^\infty \theta R \rho(x)dx.
\end{equation}
The expected degree can be written as two terms equation as

\begin{eqnarray*}
k(X) &=&  2\pi \sigma R^2
\int\limits_0^{\pi}\sin{\theta}d\theta\int\limits_{\theta R/X}^{\infty}\rho(x)dx,\\
			&=& 2\pi \sigma R^2
\left[\int_0^{\pi
R/X}\left(1-\cos{\frac{Xx}{R}}\right)\rho(x)dx+2\int_{\pi
R/X}^{\infty}\rho(x)dx\right
],\\
			&=& 2\pi 
\sigma R^2\left[\int_0^{\infty}\left(1-\cos{\frac{Xx}{R}}\right)\rho(x)dx +
\int_{\pi
R/X}^{\infty}\left(1+\cos{\frac{Xx}{R}}\right)\rho(x)dx\right],
\end{eqnarray*}
and similar equations can be written for strength and distance strength also. The second
integral can be neglected for distributions whose tail falls fast enough. In the case of
our simulations ($X_{max}=1.6$, $R=1$), the second integral is exactly $0$ since it lower
limit is $\pi/1.6\approx 2$ while the distribution $\rho(x>1.6)=0$.
After the calculation of space integrals we make an expansion of trigonometric functions
up to the
second order:
\begin{equation}
 \langle k(X)\rangle = \pi \sigma X^2 M_2(\rho) - \frac{\pi \sigma}{12 R^2} X^4 M_4(\rho)
=  \pi \sigma X^2 M_2(\rho)\left(1+c_k\right),
\end{equation}
\begin{equation}
 \langle s(X)\rangle = \pi \sigma X^3 M_3(\rho) - \frac{\pi \sigma}{12 R^2} X^5
M_5(\rho)=\pi \sigma X^3 M_3(\rho)\left(1+c_s\right),
\end{equation}
\begin{equation}
 \langle s^d(X)\rangle = \frac{2}{3}\pi \sigma X^3 M_3(\rho) - \frac{\pi \sigma}{15 R^2}
X^5 M_5(\rho)= \frac{2}{3}\pi \sigma X^3 M_3(\rho)\left(1+c_{s^d}\right).
\end{equation}
We can now calculate:
\begin{equation}
 \frac{d\langle k(X)\rangle}{dX} = 2\pi \sigma X M_2(\rho) - \frac{\pi \sigma}{3R^2} X^3
M_4(\rho)
\end{equation}
\begin{equation}
 \frac{d\langle s(X)\rangle}{dX} = 3\pi \sigma X^2 M_3(\rho) - \frac{5\pi \sigma}{12R^2}
X^4 M_5(\rho)
\end{equation}
Thus
\begin{equation}
 \frac{d\langle s\rangle}{d\langle k\rangle} = \left(3\pi \sigma X^2 M_3(\rho) -
\frac{5\pi \sigma}{12R^2} X^4 M_5(\rho)\right)\frac{1}{2\pi \sigma X M_2(\rho)}
\frac{1}{1- \frac{M_4(\rho)X^2}{6R^2M_2(\rho)}}
\end{equation}
With the assumption that corrections are small $\frac{X^2M_5(\rho)}{12R^2M_3(\rho)} \ll
1$ and $\frac{X^2M_4(\rho)}{12R^2M_2(\rho)} \ll
1 $ we get:
\begin{equation}
 \frac{d\langle s\rangle}{d\langle k\rangle} = \frac{3}{2}\frac{M_3(\rho)}{M_2(\rho)}X -
\frac{M_4(\rho)}{4R^2M_2(\rho)}\left[\frac{5}{6}\frac{M_5(\rho)}{M_4(\rho)}-
\frac{M_3(\rho)}{M_2(\rho)}\right]X^3
\end{equation}
Solving (28) for $X$ and expanding in series:
\begin{equation}
 \frac{d\langle s\rangle}{d\langle k\rangle} =
\frac{3}{2}\frac{M_3(\rho)}{\sqrt{\pi\sigma}M_2^{3/2}(\rho)}\langle k\rangle^{1/2}
+ \frac{M_4(\rho)}{48 R^2 M_2^{5/2}(\rho)
(\pi\sigma)^{3/2}}\left[15\frac{M_3(\rho)}{M_2(\rho)}-10\frac{M_5(\rho)}{M_4(\rho)}\right]
\langle k\rangle^{3/2}
\end{equation}
Integrating with condition that $\langle s\rangle = 0$ when $\langle k\rangle = 0$ we get:
\begin{equation}
 \langle s\rangle = \frac{M_3(\rho)}{\sqrt{\pi\sigma}M_2^{3/2}(\rho)}\langle
k\rangle^{3/2} 
+ \frac{M_4(\rho)}{24 R^2 M_2^{5/2}(\rho)
(\pi\sigma)^{3/2}}\left[3\frac{M_3(\rho)}{M_2(\rho)}-2\frac{M_5(\rho)}{M_4(\rho)}\right]
\langle k\rangle^{5/2}
\end{equation}
Similarly
\begin{equation}
 \frac{d\langle s^d\rangle}{d\langle k\rangle} =
\frac{M_3(\rho)}{\sqrt{\pi\sigma}M_2^{3/2}(\rho)}\langle k\rangle^{1/2}
+ \frac{M_4(\rho)}{24 R^2 M_2^{5/2}(\rho)
(\pi\sigma)^{3/2}}\left[5\frac{M_3(\rho)}{M_2(\rho)}-4\frac{M_5(\rho)}{M_4(\rho)}\right]
\langle k\rangle^{3/2}
\end{equation}
\begin{equation}
 \langle s^d\rangle = \frac{2}{3}\frac{M_3(\rho)}{\sqrt{\pi\sigma}M_2^{3/2}(\rho)}\langle
k\rangle^{3/2} 
+ \frac{M_4(\rho)}{60 R^2 M_2^{5/2}(\rho)
(\pi\sigma)^{3/2}}\left[5\frac{M_3(\rho)}{M_2(\rho)}-4\frac{M_5(\rho)}{M_4(\rho)}\right]
\langle k\rangle^{5/2}
\end{equation}
Leading order corrections to degree $c_k$ and strength $c_s$ are negligible if
$\frac{X^2M_4(\rho)}{12R^2 M_2(\rho)} \ll 1$ and $\frac{X^2M_5(\rho)}{12R^2 M_3(\rho)} \ll
1$. We calculated l.h.s. of this relations for the simulations described in this paper.
These contributions were computed for the minimal and maximal exponent of $\rho(x)$ and
for mutual couplings between ports with maximal fitness value as well as couplings between
ports with maximal fitness value and ports with typical fitness value.

The influence of the spherical geometry in the case of the sphere with homegnous
distribution of ports can be quantified via typical values of the flows i.e. products of
fitnesses:
$X_{min} = 0.1$ $X_{max} = 1.6$:\\
\begin{itemize}
 \item $X^2 = X_{max}^2$, $\gamma = 1$ : $c_k = 0.274$, $c_s = 0.328$
 \item $X^2 = X_{max} \langle X \rangle$, $\gamma = 1$ : $c_k = 0.0927$, $c_s = 0.1106$
 \item $X^2 = X_{max}^2$, $\gamma = 2.6$ : $c_k = 0.136$, $c_s = 0.2296$
 \item $X^2 = X_{max} \langle X \rangle$, $\gamma = 2.6$ : $c_k = 0.0185$, $c_s = 0.0314$
\end{itemize}

United states Airports: $X_{min} = 0.02$, $X_{max} = 0.65$:\\
\begin{itemize}
 \item $X^2 = X_{max}^2$, $\gamma = 1$ : $c_k = 7.445\cdot 10^{-3}$, $c_s = 8.926 \cdot
10^{-3}$
 \item $X^2 = X_{max} \langle X \rangle$, $\gamma = 1$ : $c_k = 2.073\cdot 10^{-3}$, $c_s
= 2.485\cdot 10^{-3}$
 \item $X^2 = X_{max}^2$, $\gamma = 2.6$ : $c_k = 3.298\cdot 10^{-3}$, $c_s = 6.172\cdot
10^{-3}$
 \item $X^2 = X_{max} \langle X \rangle$, $\gamma = 2.6$ : $c_k = 2.380\cdot 10^{-4}$,
$c_s = 4.454\cdot 10^{-4}$
\end{itemize}
 As shown, the influence of spherical geometry is practically non-existent for typical
situations and can have some effect only on ports with fitnesses drawn from the tail of
the distribution.
Spherical geometry clearly influences the relation between strength $s$, distance
strength $s^d$  and degree $k$. In general we can not expect the
infinite exponents on spherical geometries. Nevertheless, in cases
in which most of the links between ports are
below a certain threshold, the geometry is effectively almost flat and the predicted
exponents are expected to be found.

Here we calculate the expected values of degree and strength for
given $X$ in the case of finite flat geometry. Now the domain of integration is finite
and, if we assume the convexity of the domain, we can write the equation for the expected
degree of the port with fitness $X$ as:
\begin{equation}
k(X) = \int\limits_0^{2\pi}d\varphi \int\limits_0^{\Gamma(\varphi)}r dr
\int\limits_{r/X}^{\infty}\rho(x) dx,
\end{equation}
where $\Gamma(\varphi)$ describes the distance of the domain edge to the origin
with respect to angle $\varphi$.
This integral can be separated in two parts as:
\begin{eqnarray}
k(X) &= & \sigma\int\limits_0^{2\pi}d\varphi\left[ \frac{1}{2} X^2
\int\limits_0^{\Gamma(\varphi)/X}x^2\rho(x)dx +
\frac{1}{2}\Gamma^2(\varphi)\int\limits_{\Gamma(\varphi)/X}^{\infty}\rho(x)dx \right],\\
		&=&\pi \sigma X^2 \int\limits_0^{\infty}x^2\rho(x)dx - \frac{1}{2} \sigma
X^2\int\limits_0^{2\pi}d\varphi\int\limits_{\Gamma(\varphi)/X}^{\infty}\left[1-\left(\frac
{\Gamma(\varphi)}{Xx}\right)^2\right]x^2\rho(x)dx.
\end{eqnarray}
For the strength the equation is very similar:

\begin{equation}
s(X) = \pi \sigma X^3 \int\limits_0^{\infty}x^3\rho(x)dx - \frac{1}{2} \sigma
X^3\int\limits_0^{2\pi}d\varphi\int\limits_{\Gamma(\varphi)/X}^{\infty}\left[1-\left(\frac
{\Gamma(\varphi)}{Xx}\right)^2\right]x^3\rho(x)dx.
\end{equation}
These expressions can be rewritten in a more transparent form as:

\begin{eqnarray}
k(X) &=&\pi \sigma X^2 \int\limits_0^{\infty}x^2\rho(x)dx\left[1 -
\frac{1}{2\pi}\int\limits_0^{2\pi}d\varphi\frac{\int\limits_{\Gamma(\varphi)/X}^{\infty}
\left[1-\left(\frac{\Gamma(\varphi)}{Xx}\right)^2\right]x^2\rho(x)dx}{\int\limits_0^{
\infty}x^2\rho(x)dx}\right]\\
		&=& \pi \sigma X^2 \int\limits_0^{\infty}x^2\rho(x)dx \left[1 -
\left\langle
\epsilon_k\left(\frac{\Gamma(\varphi)}{X}\right) \right\rangle_{\varphi}\right],
\end{eqnarray}
and

\begin{eqnarray}
s(X) &=&\pi \sigma X^3 \int\limits_0^{\infty}x^3\rho(x)dx\left[1 -
\frac{1}{2\pi}\int\limits_0^{2\pi}d\varphi\frac{\int\limits_{\Gamma(\varphi)/X}^{\infty}
\left[1-\left(\frac{\Gamma(\varphi)}{Xx}\right)^2\right]x^3\rho(x)dx}{\int\limits_0^{
\infty}x^3\rho(x)dx}\right]\\
		&=& \pi \sigma X^3 \int\limits_0^{\infty}x^3\rho(x)dx \left[1 -
\left\langle
\epsilon_s\left(\frac{\Gamma(\varphi)}{X}\right) \right\rangle_{\varphi}\right].
\end{eqnarray}
If we assume an upper cut-off of the distribution $X_{max}$, an assumption which is very
realistic for any imaginable port system, then if $\forall
\varphi:\Gamma(\varphi) > XX_{max}$ both $\epsilon$ functions are equal to zero. 
If this condition is not fulfilled, a weaker condition will still preserve infinite limit
exponents. Namely, a factor $1 -
\left(\frac{\Gamma(\varphi)}{Xx}\right)^2$ can achieve values from $0$ for $x =
\frac{\Gamma(\varphi)}{X}$ up to $1$ for large $X$ which reduces the contribution of
the second term. Finally, by averaging over the angle $\varphi$ with the assumption of
a convex domain the contribution of terms different than zero is reduced by factor $2$ at
worst.
Finally, if we assume $X_{max} \leq
\sqrt{\Gamma}$ where $\Gamma$ represents the characteristic linear
dimension of the domain, we can easily expect the infinite limit exponents recovered too. 

\newpage

\section*{Supplementary information 4}

In the hypothetical case of a D-dimensional infinite plane, our model is integrated over
D-dimensional volume with the differential element $dV_D=r^{D-1}C_Ddr$. From this
trivially
follows 

\begin{eqnarray}
\langle k(X)\rangle &=& \sigma
C_D\int_0^\infty r^{D-1}dr\int_{0}^{\infty}\Theta\left(xX-r\right)\rho(x)dx,\\
\langle s(X)\rangle &=& \sigma
C_D\int_0^\infty r^{D-1}dr\int_{0}^{\infty}xX\Theta\left(xX-r\right)\rho(x)dx,\\
\langle s^d(X)\rangle &=& \sigma
C_D\int_0^\infty r^{D-1}dr\int_{0}^{\infty}r\Theta\left(xX-r\right)\rho(x)dx.\\
\end{eqnarray}
One can see that the exponents $\beta=\beta^d=(D+1)/D$ and $\alpha=1/D$

\newpage

\begin{figure}
\includegraphics[width=0.88\textwidth]{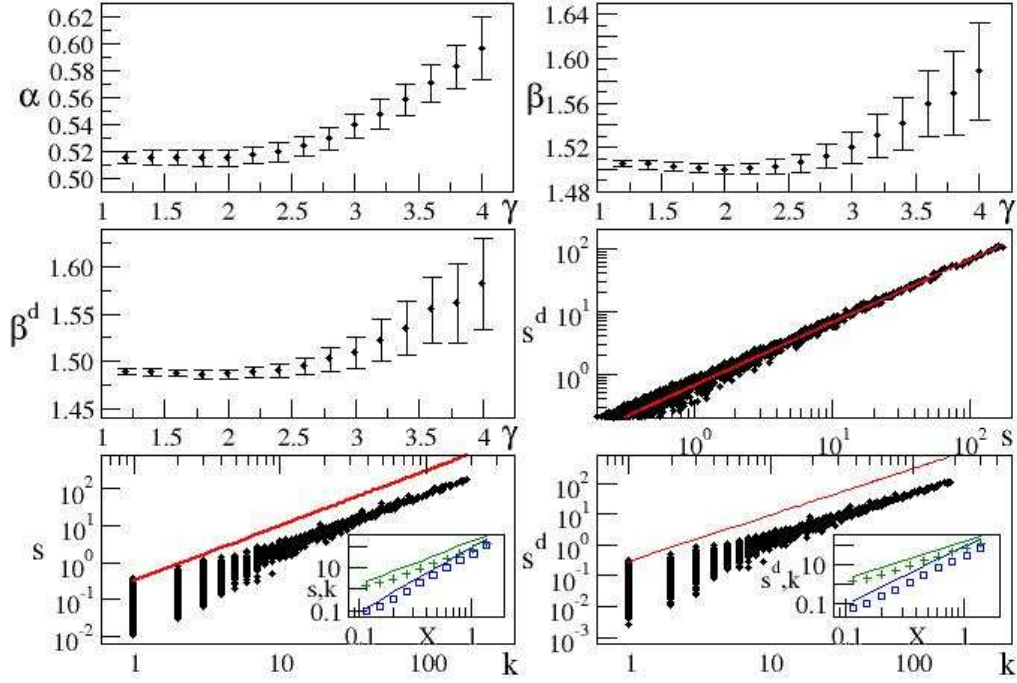}
\caption{\label{Figure1} Simulations for a finite 2D plane are represented in these
figures. The upper left figure represents the $\alpha$
dependence on the power law exponent $\gamma$, the upper right figure represents the
$\beta$
dependence on the power law exponent $\gamma$, and the middle left figure represent the
$\beta^d$ dependence on the power law exponent $\gamma$. It is clear that infinite
exponents are well reproduced in finite flat space as well. The middle right figure
represents
the relation between strength and distance strength in the simulated data. The red line is
the theoretically predicted result of $s=2/3s^d$. The lower left figure and the lower
right
figure
represent the strength $s$ and degree strength $s^d$ dependence on degree $k$. The data
points
are simulations and the red line has an exponent $3/2$ and is drawn for comparison. 
Strength $s$, distance strength $s^d$ and degree $k$ as a function of fitness
$X$ are in the insets. The data in insets are exponentially binned to improve visibility
and lines with
exponents $3$ and $2$ are drawn for comparison. The networks for this simulation are
realized on a finite flat space as described in the Methods section. The exponents are
well
reproduced for a broad range of distribution parameters.}
\end{figure}

\begin{figure}
\includegraphics[width=0.88\textwidth]{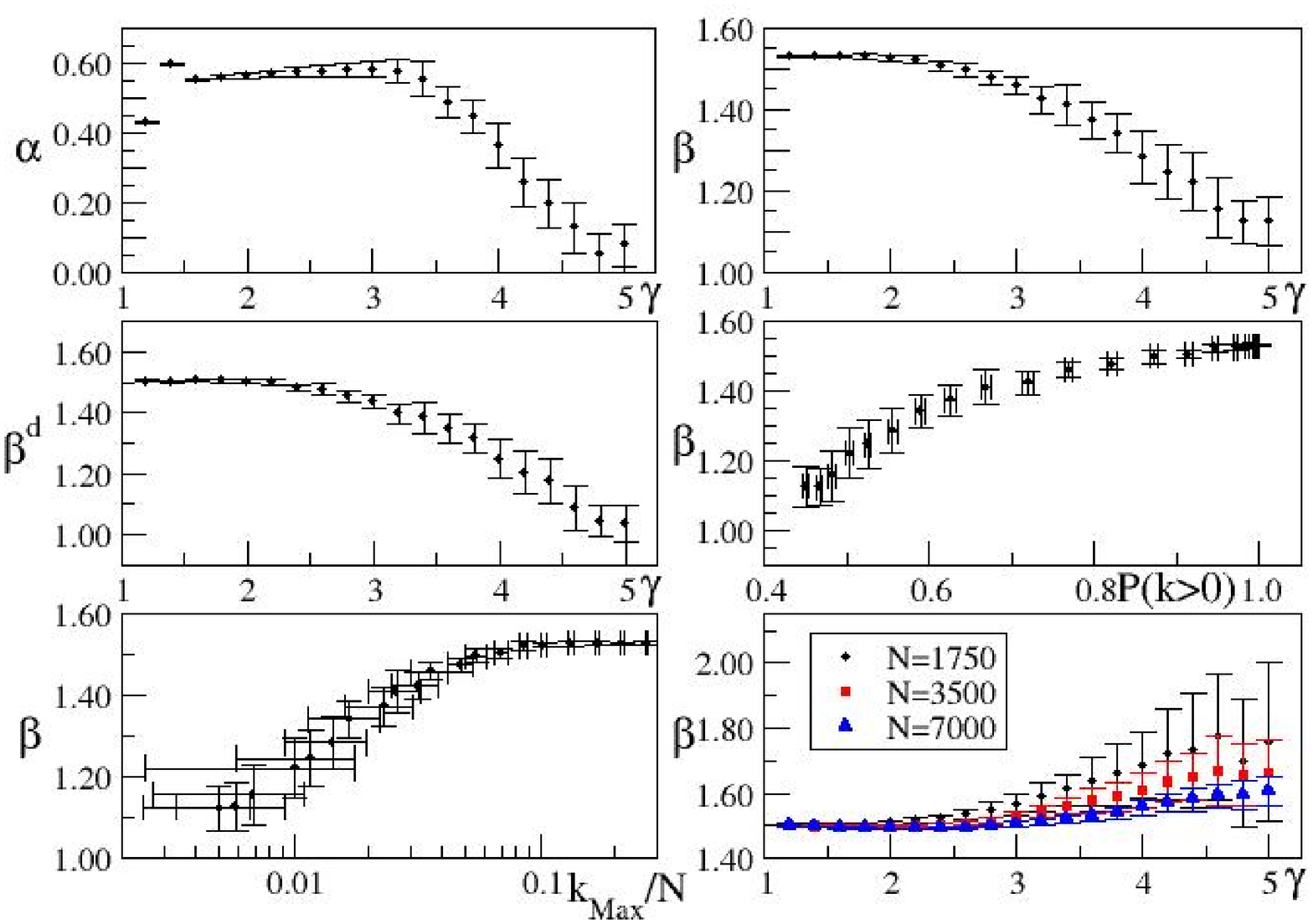}
\caption{\label{Figure2} The upper left figure represents the $\alpha$ dependence on the
power law exponent $\gamma$, the upper right figure represents the $\beta$ dependence on
the power law exponent $\gamma$, and the middle left figure represent the $\beta^d$
dependence on the power law exponent $\gamma$. Locations of points are locations of
$18745$\cite{GeoLocations} American airports and fitness variables are randomly drawn from
the power law distributions described with exponent $\gamma$, minimal value of fitness
$x_{min}=0.02$ and maximal value of fitness $x_{max}=0.65$. It is clear that infinite
exponents are well reproduced in spherical geometry for $\gamma<3$. The middle right
figure represents the dependence of $\beta$ on the fraction of airports $P(k>0)$ which
have at
least one neighbour. It is clear that the simulated exponents differ from infinite
ones when the network starts to disconnect due to a big number of airports with small
fitness
parameters. The lower left figure represents the relationship between $\beta$ and
the maximal
degree of the airport in the network. Again, when the maximal degree is too small,
networks
effectively disconnect and simulated exponents deviate from the infinite case. The lower
right figure demonstrates that higher density of points reproduces the infinite
exponents better. The relationship between $\beta$ and $\gamma$ for three cases
of the different density of points in the case of a sphere of radius $R=1$ is presented.
$N$ is the number of points on the sphere and the density is $\sigma=N/4\pi$.   }
\end{figure}

\begin{figure}
\includegraphics[width=0.88\textwidth]{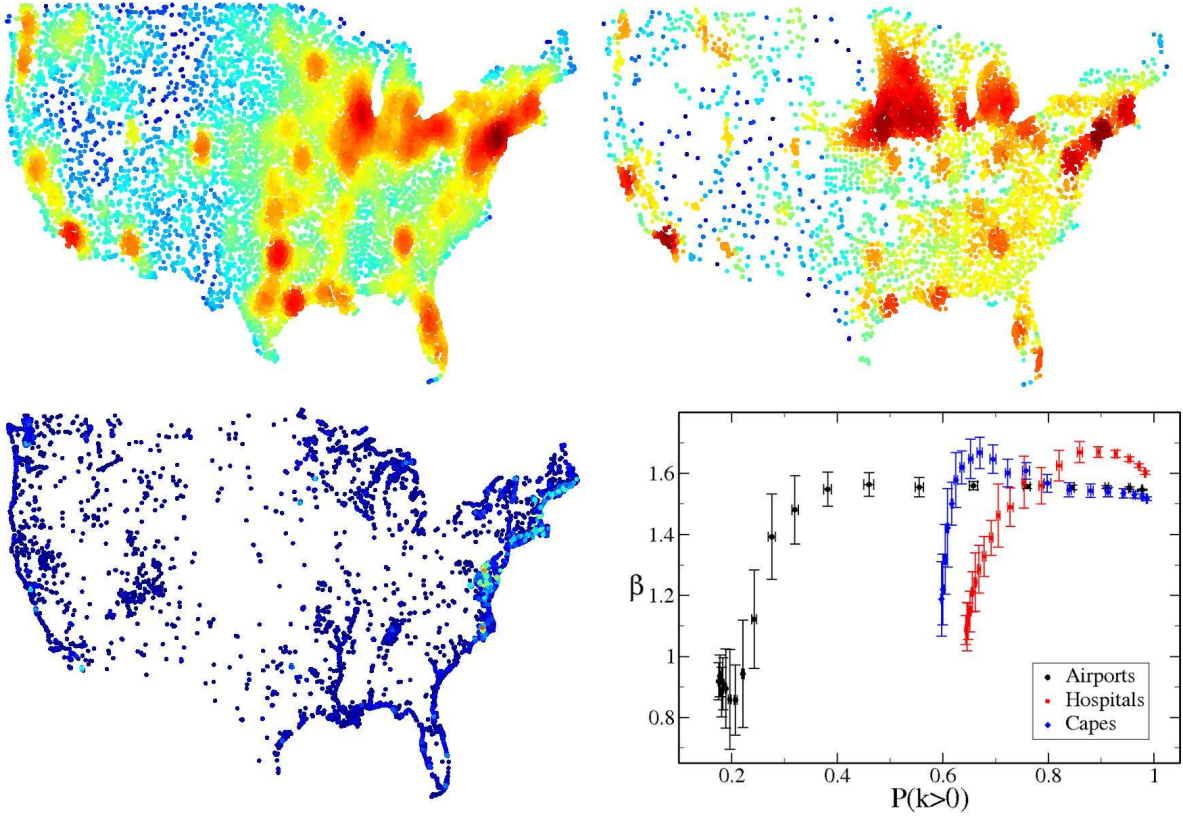}
\caption{\label{Figure3} The upper left figure represents locations of $18745$ American
airports, the upper right figure represents locations of $13385$ American hospitals and
the lower
left figure represents locations of $11938$ American capes. Colours represent different
local densities of locations in order to emphasize differences among heterogeneities of
this data. The relationship between $\beta$ and $P(k>0)$ for these sets of ports is shown
in the lower right figure. The minimal value of fitness is
$x_{min}=0.01$ and maximal value of fitness is $x_{max}=0.65$. The infinite exponent is
again reproduced for all three sets of spatial points, as long as $\rho(x)$ is
sufficiently broad i.e. a network is connected.  }
\end{figure}

\begin{figure}
\includegraphics[width=0.48\textwidth]{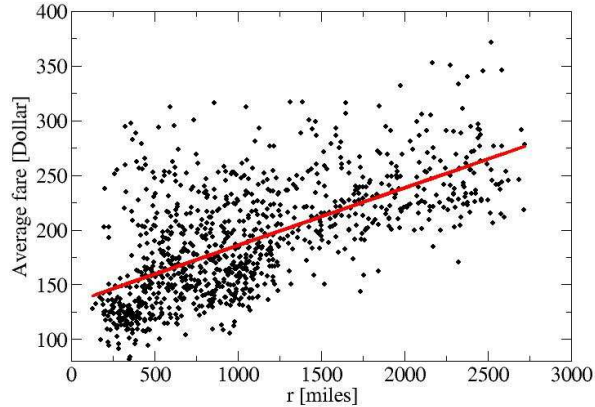}
\caption{\label{AirFares} The average fares of travel among 200 largest american
airports are placed on the y-axis, and on the x-axis is the mutual distance of airports.
Data points represent
the average
fare of travel in the third quarter of 2010 between two airports at a given distance.
The data was
downloaded from \cite{AirFareData}. The straight line represents linear regression
obtained via least
squares. Pearson's correlation of the data is $0.64$. It is visible that the smallest
fares
at a given distance also have a linear like behaviour, and we can expect that the
smallest
fares are a better representation of distance costs then the average fares.}
\end{figure}

\begin{figure}
\includegraphics[width=0.48\textwidth]{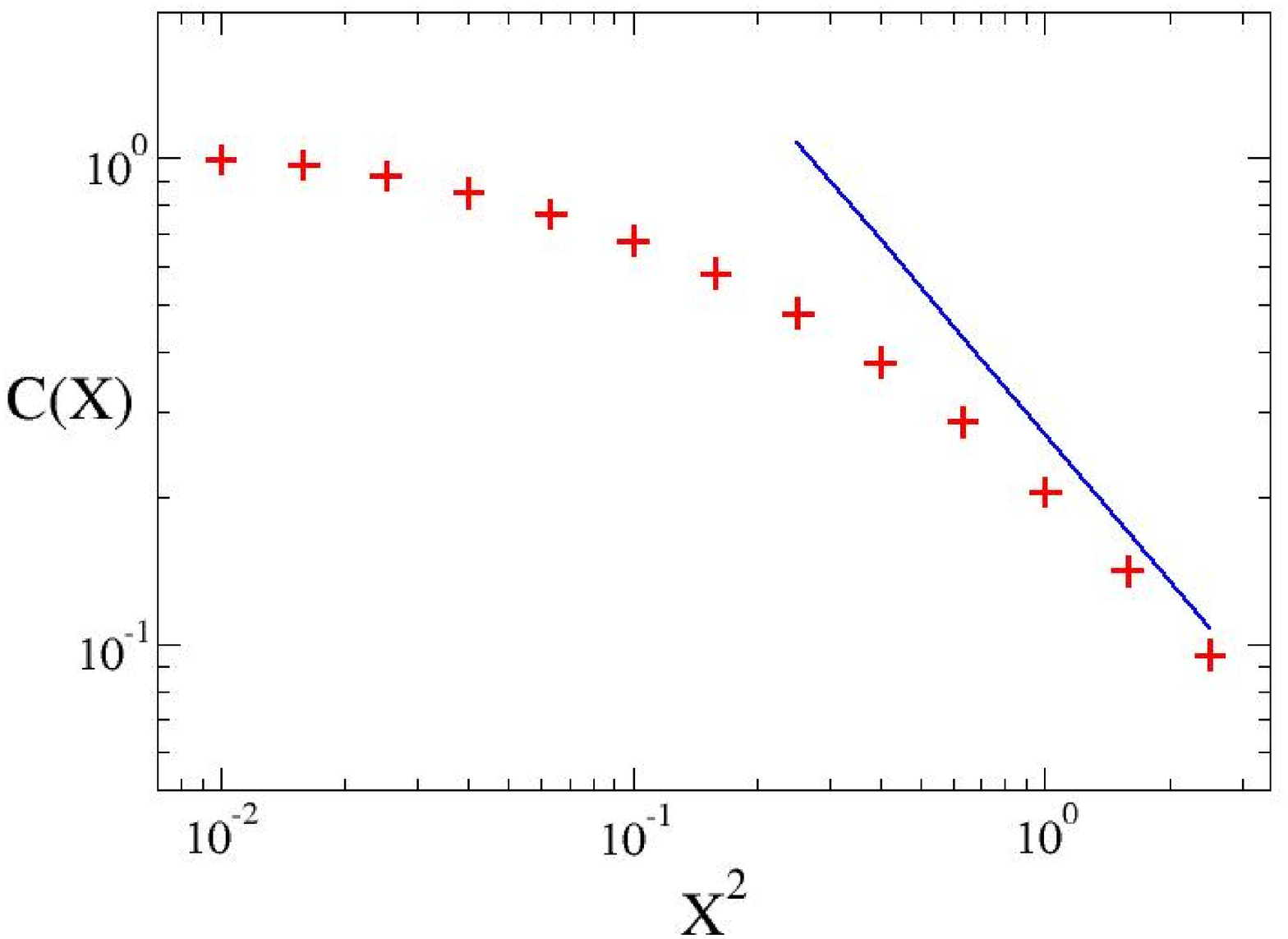}
\caption{\label{Clustering} An average value of the clustering coefficient $C$ is placed
Onon the y-axis, and the square of the fitness $X^2$ is placed on the x-axis. The data
points (pluses) represent
the expected clustering dependent on the square of $x^2\sim k$. The full line represents
the
power law with exponent $-1$ and such assymptotic relationship is found in real networks.
The minimal and maximal value of $x$ are again $0.1$ and $1.6$, which were used in
simulations
as well.}
\end{figure}

\newpage


\begin{thebibliography}{1}

\bibitem{Guimera2005}Guimer\'a, R., Mossa, S., Turtschi, A. \& Amaral,L. A. N. 
The worldwide air transportation network: Anomalous
centrality, community structure, and cities global roles.
\emph{Proc. Natl. Acad. Sci.}, \textbf{102}, 7794–7799 (2005).

\bibitem{Colizza2006}
Colizza, V., Barrat, A., Barthélemy, M., Vespignani, A. The role of the airline
transportation network in the prediction and predictability of global epidemics. 
\emph{Proc. Natl. Acad. Sci.}, \textbf{103}, 2015–2020 (2006).

\bibitem{Imo2006}
International Maritime Organization. International shipping and world trade. Facts and
figures. Available at \url{http://www.imo.org/}, (2006).

\bibitem{Ruiz2000}
Ruiz, G. M., Rawlings, T. K., Dobbs, F. C., Drake, L. A., Mullady, T., Huq, A. \&
Colwell, R. R.
Global spread of microorganisms by ships. \emph{Nature}, \textbf{408}, 49–50 (2000).

\bibitem{Barthelemy2010} Barthelemy, M. Spatial Networks,
\url{http://arxiv.org/abs/1010.0302}

\bibitem{Barrat2004} Barrat, A., Barth\'elemy, M., Pastor-Satorras, R. \& Vespignani, A.
The architecture of complex weighted networks.
\emph{Proc. Natl. Acad. Sci.}, \textbf{101}, 3747--3752 (2004).

\bibitem{Barrat2005}Barrat, A., Barth\'elemy, M., \& Vespignani, A. The effects of spatial
constraints on the evolution of weighted complex networks, \emph{J. Stat. Mech.}, P05003
(2005).                      ́

\bibitem{Bagler2008} Bagler,G., Analysis of Airport Network of India as a Complex
Weighted Network, \emph{Physica A}, \textbf{387}, Issue 12, 2972--2980 (2008). 

\bibitem{Zhang2010} Zhang, J., Cao, X-B., Du, W-B.,\and Cai, K-Q. 
,Evolution of Chinese airport network,\emph{Physica A}, \textbf{389}, 3922--3931 (2010). 

\bibitem{Brockman} O'Danleyman, G, Lee, J.J., Seebens, H., Blasius, B. ,\and Brockman,
D., Complexity in human transportation networks: A comparative analysis of worldwide
air transportation and global cargo ship movements,\url{arXiv:/1103.5451}.

\bibitem{Kaluza2010} Kaluza, P., Koelzsch, A., Gastner, M.T., \& Blasius, B. The complex
network of global cargo ship movements. \emph{J. R. Soc. Interface}, \textbf{7}, 1093 –
1103 (2010).

\bibitem{Hu2009}
Hu, Y. \& Zhu, D. Empirical analysis of the worldwide maritime transportation network.
\emph{Physica A}, \textbf{388}, 2061 -- 2071 (2009).


\bibitem{Li2004} Li, W. \& Cai, X. Statistical analysis of airport network of china.
\emph{Phys. Rev. E}, \textbf{69}, 046106 (2004).

\bibitem{Guimera2004}  Guimer\'a, R. \& Amaral,L. A. N. Modeling the world-wide airport
network. \emph{Eur. Phys. J. B}, \textbf{38}, 381 – 385 (2004).

\bibitem{Xu2007} Xu, X., Hu, J. \& Liu, P. Empirical analysis of the ship-transport
network of china. \emph{Chaos}, \textbf{17}, 023129 (2007).

\bibitem{Mukherjee2006} Mukherjee, G. \& Manna, S. S. Weighted scale-free networks in
Euclidean space using local selection rule. \emph{Phys. Rev. E}, \textbf{74},
036111 ͑(2006͒).

\bibitem{Barabasi1999}
Barab\'asi A.-L. \& Albert, R.,Emergence of Scaling in Random Networks. \emph{Science},
\textbf{286}, 509 ͑(1999͒).

\bibitem{Barrat2004Prl}Barrat, A., Barth\'elemy, M., \& Vespignani, A.
Weighted Evolving Networks: Coupling Topology and Weight Dynamics. \emph{Phys. Rev.
Lett.}, \textbf{92}, 228701 (2004).

\bibitem{Ginestra2005} Bianconi, G. Emergence of weight-topology correlations in complex
scale-free networks. \emph{Europhys. Lett.}, \textbf{71}, 1029 (2005).

\bibitem{Caldarelli2002}
Caldarelli, G., Capocci, A., De Los Rios, P., and Mun\~{o}z, M. A. Scale
 free networks from varying vertex intrinsic fitness. \emph{Physical Review Letters},
\textbf{89}, 258702 (2002).

\bibitem{Goldenfled1992}
Goldenfeld, N. D.  Lectures on Phase Transitions and the Renormalisation Group
\emph{Addison-Wesley}, (1992).

\bibitem{Newman2003}Newman, M.E.J., Mixing patterns in networks. \emph{Phys. Rev. E},
\textbf{67}, 026126 (2003).

\bibitem{AirFareData}\url{
http://ostpxweb.dot.gov/aviation/X-50\%20Role_files/consumerairfarereport.htm}

\bibitem{CaldarelliBook} Caldarelli, G., \emph{Scale-Free Networks}, Oxford University
Press, Oxford, (2007).

\bibitem{Feller}
Feller, W., \emph{An Introduction to Probability theory and its applications}, 
John Wiley \& sons, New York London Sydney, (1968).

\bibitem{GeoLocations}\url{http://geonames.usgs.gov/domestic/download_data.htm}

\bibitem{West1997} West, G.B., Brown, J.H., and Enquist, B.J., A General Model for the
Origin of Allometric Scaling Laws in Biology, \emph{Science}, \textbf{276} (5309),
122-126 (4 April 1997).
\end{thebibliography}
\end{document}